\documentclass[aps,twocolumn,prx,superscriptaddress,longbibliography]{revtex4-2}

\usepackage{graphicx}
\usepackage{dcolumn}
\usepackage{bm}
\usepackage[T1]{fontenc}
\usepackage[latin9]{inputenc}
\usepackage{textcomp}
\usepackage{amsmath}
\usepackage{mathrsfs}
\usepackage{bbold}
\usepackage{physics}
\usepackage{amssymb}
\usepackage{units}
\usepackage{ulem}
\usepackage{xcolor}
\usepackage{comment}
\usepackage[colorlinks,bookmarks=false,citecolor=blue,linkcolor=blue,urlcolor=blue]{hyperref}
\usepackage{soul}

\begin{document}

\title{Thouless pumping in a driven-dissipative Kerr resonator array}

\author{S. Ravets}
\email[]{sylvain.ravets@c2n.upsaclay.fr}
\affiliation{Universit\'{e} Paris-Saclay, CNRS, Centre de Nanosciences et de Nanotechnologies (C2N),
91120 Palaiseau, France}

\author{N. Pernet}
\affiliation{Universit\'{e} Paris-Saclay, CNRS, Centre de Nanosciences et de Nanotechnologies (C2N),
91120 Palaiseau, France}

\author{N. Mostaan}
\affiliation{Department of Physics and Arnold Sommerfeld Center for Theoretical Physics (ASC), Ludwig-Maximilians-Universit\"at M\"unchen, Theresienstr. 37, D-80333 M\"unchen, Germany}
\affiliation{Munich Center for Quantum Science and Technology (MCQST), Schellingstr. 4, D-80799 M\"unchen, Germany}
\affiliation{CENOLI,
Universit\'e Libre de Bruxelles, CP 231, Campus Plaine, B-1050 Brussels, Belgium}

\author{N. Goldman}
\affiliation{CENOLI,
Universit\'e Libre de Bruxelles, CP 231, Campus Plaine, B-1050 Brussels, Belgium}
\affiliation{Laboratoire Kastler Brossel, Coll\`ege de France, CNRS, ENS-PSL University, Sorbonne Universit\'e, 11 Place Marcelin Berthelot, 75005 Paris, France}

\author{J. Bloch}
\affiliation{Universit\'{e} Paris-Saclay, CNRS, Centre de Nanosciences et de Nanotechnologies (C2N),
91120 Palaiseau, France}

\begin{abstract}

Thouless pumping is an emblematic manifestation of topology in physics, referring to the ability to induce a quantized transport of charge across a system by simply varying one of its parameters periodically in time. The original concept of Thouless pumping involves a non-interacting system, and has been implemented in several platforms. One current challenge in the field is to extend this concept to interacting systems. In this article, we propose a Thouless pump that solely relies on nonlinear physics, within a chain of coupled Kerr resonators. Leveraging the driven-dissipative nature of the system, we modulate in space and time the onsite Kerr interaction energies, and generate 1+1-dimensional topological bands in the Bogoliubov spectrum of excitations. These bands present the same topology as the ones obtained within the Harper-Hofstadter framework, and the Wannier states associated to each band experience a net displacement and show quantized transport according to the band Chern numbers. Remarkably, we find driving configurations leading to band inversion, revealing an interaction-induced topological transition. Our numerical simulations are performed using realistic parameters inspired from exciton polaritons, which form a platform of choice for investigating driven topological phases of matter.

\end{abstract}

\maketitle

A one-dimensional system of non interacting electrons in a time-modulated periodic potential can show transport in absence of any external bias, a phenomenon known as geometric pumping. Strikingly, there exists modulation protocols where the electron motion is quantized in direct connection to the topology of the 1+1-dimensional (1+1D) bands that the system outlines during the protocol, in analogy to the quantum Hall effet~\cite{Thouless1983}. In the case of filled bands, electrons move during a pump period by an integer number of unit cells equal to the sum of the Chern numbers of the filled bands~\cite{Citro2023}. Noteworthily, as quantized transport is intrinsically linked to the band structure, it can be accessed by monitoring the motion of the Wannier states. This way, Thouless pumping can also be accessed in bosonic systems. 

The development of synthetic experimental platforms, providing exquisite control over the system parameters, has enabled experimental demonstrations of Thouless pumping. Examples include cold atoms confined in tunable optical lattices~\cite{Lohse2016,Nakajima2016,Lu2016} or coupled to a cavity field~\cite{Dreon2022}, photons in coupled waveguides~\cite{Kraus2012,ke2016,Cerjan2020}, plasmons in plasmonic waveguides \cite{Fedorova2020}, or mechanical excitations in metamaterials~\cite{Grinberg2020} or waveguides~\cite{Xia2021}. Generalizations to non-Abelian (degenerate-band) settings were experimentally explored in acoustic or optical waveguides~\cite{You2022,Sun2022}.
 
The picture gets even richer when considering interacting particles~\cite{Niu1984}. Interactions can induce topological pumping \cite{Kuno2020,viebahn2023}, or lead to its breakdown \cite{Tuloup2023,Walter2023} depending on the interaction strength. Nonlinear solitonic waves on top of a Thouless pump have shown to exhibit quantized transport \cite{Jurgensen2021,jurgensen2022chern,Mostaan2022,Fu2022,Jurgensen2023,Cao2024}. Thouless pumping has been proposed as a way to channel multiphoton quantum states~\cite{Ke2017,Tangpanitanon2016}. Remarkably, interactions in bosonic systems are also known to renormalize the band structure for the elementary Bogoliubov excitations~\cite{Bogoliubov1947}, and have been shown to induce non trivial topology in the Bogoliubov bands of an otherwise trivial system~\cite{Bardyn2016,DiLiberto2016,Bleu2016,Septembre2023}. To the best of our knowledge, the question whether one can organize a Thouless pump for the Bogoliubov modes of an interacting bosonic system remains unexplored so far. 

In this Letter, we propose an optically driven topological pump that entirely relies on the optical nonlinearity in a 1D array of driven-dissipative Kerr resonators. We consider a protocol where an optical drive is periodically distributed across the lattice sites with adiabatic time-periodic intensity modulation. Because of the Kerr nonlinearity, this protocol induces a modulation of the lattice onsite energies, proportionally to the local photon number. As a result, topological bands akin to Harper-Hofstadter bands emerge in the 1+1D Bogoliubov spectrum of excitations and Thouless pumping of the Wannier states is demonstrated. Moreover, we show that anaharmonicities in the modulation protocol may induce topological transitions. We anticipate that this protocol could be implemented in a wide variety of experimental platforms such as coupled photonic resonators, superconducting circuits, cold atoms in cavities, opto-mechanical resonators or micro-electromechanical resonators~\cite{Smirnova2020}. In this Letter, we base our numerical simulations on realistic parameters that are relevant for the exciton polariton lattices~\cite{Carusotto2013,Schneider2017}.

\begin{figure}
    \centering
    \includegraphics{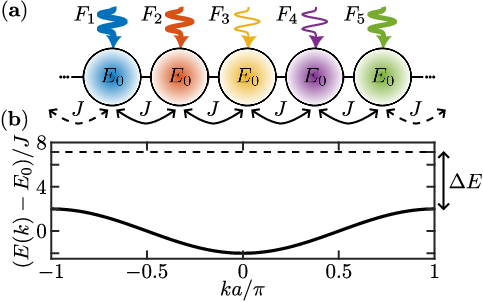}
    \caption{a.~Schematic of the lattice of coupled Kerr resonators. Each lattice site is individually driven (zigzag arrows). b.~Energy dispersion of the lattice modes in the linear regime. The horizontal dashed line shows the energy $\hbar \omega_{\rm P}$ of the drives.}
    \label{Fig1}
\end{figure}

We consider an infinite periodic chain of identical Kerr resonators with constant couplings along the chain (see Fig.~\ref{Fig1}a). The linear part of the Hamiltonian writes:

\begin{equation}
\begin{aligned}
\hat{H} = \sum_{n} E_0 \dyad{\psi _n}{\psi _n} - J \left( \dyad{\psi _n}{\psi _{n+1}} + \dyad{\psi _n}{\psi _{n-1}} \right)  \ ,
\end{aligned}
\label{Eq:Hamiltonian}
\end{equation}

\noindent where $\left| \textcolor{purple}{\psi} _n \right>$ is the mode of resonator $n$ with on-site energy $E_0$, and $J$ is the coupling between neighboring sites. The eigenenergies of the Bloch eigenstates form a single band of dispersion $E(k) = E_0 - 2 J \cos{(k a)}$, where $a$ is the lattice period and $k \in ]-\pi /a, \pi / a]$ (see Fig.~\ref{Fig1}b). The nonlinear resonators are coherently driven by a monochromatic field oscillating at angular frequency $\omega_{\rm  p}$. The complex field amplitude at site $n$ is $F_n$, and all sites experience a uniform loss rate $\gamma$. Due to the driven-dissipative nature of the problem, the full quantum description of the system can be formulated in terms of a quantum master equation for a bosonic setting with a coherent drive. Here, we focus on the mean-field regime, where this equation reduces to the driven-dissipative Gross Pitaevskii equation, as was derived in ~Ref.~\cite{Carusotto2013}. Here, we use a discrete version of this equation to describe the amplitude dynamics of the field on each lattice site:

\begin{equation}
\begin{aligned}
    i \hbar \partial_t \boldsymbol{\psi} = & \left( \hat{H}    -i\frac{\gamma}{2}  \hat{\mathbb{1}}  + g  \hat{N} \right) \boldsymbol{\psi}  + i \boldsymbol{F} e^{- i \omega_{\rm p} t} \, ,
\end{aligned}
\label{Eq:GPE}
\end{equation}

\noindent where $g$ is the Kerr nonlinearity, $\boldsymbol{\psi}$ is a vector of components $\psi_n$, $\hat{N}$ is a diagonal operator with non-zero elements equal to $\hat{N}_{n,n} = |\psi_n|^2$, $\hat{\mathbb{1}}$ is the identity matrix, and $\boldsymbol{F}$ is a vector of components $F_n$. We define the energy detuning between the drive and the top of the band $\Delta E = \hbar \omega_{\rm P} - E_0 - 2J$ (see Fig.~\ref{Fig1}b), and the drive power $P = \sum_{n} \, \left| F_n \right|^2$.

We propose a drive protocol where we spatially modulate the pump amplitudes as follows: 
\begin{equation}
F_{n} ( \varphi ) = F \left| \cos (\varphi/2 + \pi \alpha \, n   ) \right| \, .
\label{Eq:protocol}
\end{equation}
\noindent The parameter $\varphi$ is a phason that can take values between $0$ and $2 \pi$, and $\alpha$ is a real parameter that encodes the spatial pattern of the drive. For rational values $\alpha=p/q$, the protocol is spatially periodic, with $q$ sites per unit cell, while irrational values of $\alpha$ lead to a quasi-periodic drive pattern. In the main text, we discuss periodic drive protocols, and choose $\alpha=1/5$, without loss of generality. The spatial periodicity of this protocol is thus $5 a$, with five sites per unit cell. Additional data for two other values of $\alpha$ can be found in the Supplemental~\cite{SupMat}.

\begin{figure}
    \centering
    \includegraphics{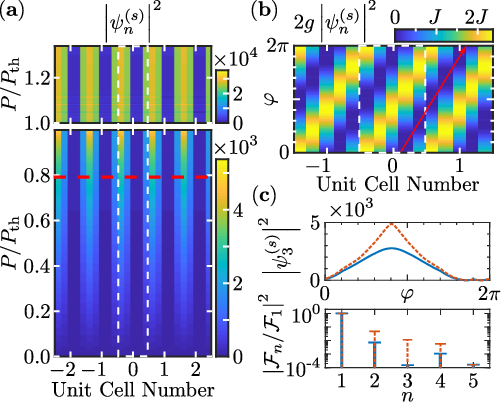}
    \caption{a. Stationary intensity map versus unit cell number and power for the lattice driven according to $\varphi = 0$ in Eq.~\ref{Eq:protocol}. The dotted horizontal line marks the drive power $P = 0.79 P_{\rm th}$ used in most part of this letter. b.~Nonlinear onsite energies $2 g \left| \psi_n ^{(s)} (F_n(\varphi)) \right|^2$ as a function of $\varphi$. The red arrow is a guide to the eye to highlight the trajectory for one energy minimum. The vertical white dashed lines in a. and b. delimit the five sites forming one unit cell. c)~Top panel: the population $\left| \psi_3 ^{(s)} \right|^2$ on site 3 of the unit cell is shown as a function of $\varphi$ for $P/P_{\rm th} = 0.79$ (solid blue line) and for $P/P_{\rm th} = 0.996$ (dashed red line). The associated power spectra (normalized to the power in the fundamental harmonic $\left| \mathcal{F}_1 \right| ^2$) are plotted in the bottom panel. The numerical parameters are $\gamma = 44.7~{\rm \mu eV}, \, J/ \gamma = 2.2, \, g/\gamma = 10 ^{-3}, \, \Delta E = 5.14 J$.}
    \label{Fig2}
\end{figure}

We first set $\varphi = 0$ and compute the system nonlinear steady-state versus $P$, solving Eq.~\ref{Eq:GPE}. We plot, in Fig.~\ref{Fig2}a, the stationary field intensity pattern $ \left| \psi_n^{(s)} \right| ^2$ calculated for increasing values of $P$. The intensity pattern follows the periodicity of the driving field, and evidences a nonlinear threshold at $P = P_{\rm th}$, characteristic of a multi-stable nonlinear system. In the following, unless stated otherwise, we set $P/P_{\rm th} = 0.79$ (red dashed line in Fig.~\ref{Fig2}a).

We now compute the excitation spectrum on top of this nonlinear steady state. To do so, we use a Bogoliubov approach writing: $ \boldsymbol{\psi} (t) = (\boldsymbol{\psi} ^{(s)} + \delta \boldsymbol{\psi} (t)) e^{-i \omega_{\rm P}t}$.  
Introducing this ansatz into the Gross-Pitaevskii equation, we obtain the following linearized equations for $\delta \boldsymbol{\psi} (t)$ and $\delta \boldsymbol{\psi}^* (t)$:

\begin{equation}
\begin{aligned}
i \hbar \partial_t
\begin{pmatrix}
\delta \boldsymbol{\psi} (t)\\
\delta \boldsymbol{\psi}^* (t)
\end{pmatrix}
=
\begin{pmatrix}
\hat{\mathcal{M}} & \hat{\mathcal{N}} \\
-\hat{\mathcal{N}}^{*} & -\hat{\mathcal{M}}^{*}
\end{pmatrix}
\begin{pmatrix}
\delta \boldsymbol{\psi} (t)\\
\delta \boldsymbol{\psi}^* (t)
\end{pmatrix} \, ,
\end{aligned}
\label{Eq:Bogo}
\end{equation}

\noindent where $\hat{\mathcal{M}} \! = \! \hat{H} - \big( \hbar \omega_{P}+ i \frac{\gamma}{2} \big) \hat{\mathbb{1}} + 2 g  \hat{N}^{(s)}$, and $\hat{\mathcal{N}}$ is a diagonal operator with non-zero elements equal to $\hat{\mathcal{N}}_{n,n} = g \left( \psi_n ^ {(s)} \right)^2$.

The stationary solutions of Eq.~\ref{Eq:Bogo} have the form: $\delta \boldsymbol{\psi} (t)= \boldsymbol{u} e^{ - i \mathcal{E} t / \hbar} + \boldsymbol{v}^{*} e^{i \mathcal{E} ^{*} t / \hbar}$, where $(\boldsymbol{u} , \boldsymbol{v}) ^ { \mathrm{T} }$ are eigenvectors of the Bogoliubov matrix written in Eq.~\ref{Eq:Bogo} with eigenvalues $\mathcal{E}$. In the Bogoliubov matrix, $\hat{\mathcal{M}}$ (resp. $-\hat{\mathcal{M}^*}$) is associated to the normal (ghost) modes, while $\hat{\mathcal{N}}$ describes the coupling between the normal and ghost modes. The set of all possible eigenvalues $\operatorname{Re}(\mathcal{E})$ constitutes the Bogoliubov spectrum of excitations. It is convenient to write the Bogoliubov eigenmodes in reciprocal space, using the basis of Bloch states $(\boldsymbol{\Tilde{u}}(k), \boldsymbol{\Tilde{v}}(k)) ^ {\mathrm{T}}$, where $k$ denotes quasi-momentum. The components $\boldsymbol{\Tilde{u}}_i(k)$ and $\boldsymbol{\Tilde{v}}_i(k)$ then denote the ``normal'' and ``ghost''  amplitudes, in band $i$ at quasi-momentum $k$.

\begin{figure}
    \centering
    \includegraphics{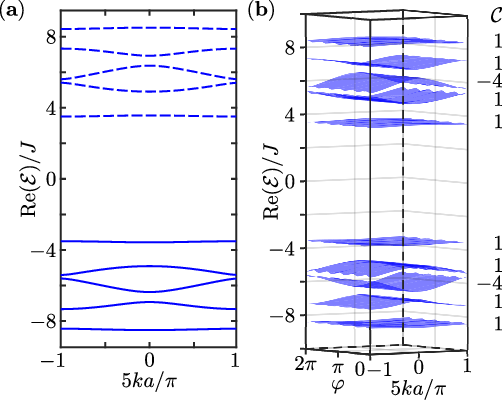}
    \caption{a.~Bogoliubov bands on top of the nonlinear steady-state $\psi^{(s)}$. The normal (ghost) branches are shown in solid (dashed) lines. b.~1+1D Bogoliubov bands as a function of $k$ and $\varphi$. The computed Chern numbers are given aside each band. $\gamma = 44.7~{\rm \mu eV}, \, J/ \gamma = 2.2, \, g/\gamma = 10 ^{-3}, \, \Delta E = 5.14 J, \, P/P_{\rm th}=0.79$}.
    \label{Fig3}
\end{figure}

Figure~\ref{Fig3}a shows the resulting Bogoliubov spectrum versus $k$. Because of the $5a$ spatial period imposed by the drive, $k$ takes values within $ ]-\pi /5a, \pi / 5a]$. We clearly observe the presence of 10 bands separated by energy gaps. By construction, the bands are symmetric with respect to the pump energy $\operatorname{Re}(\mathcal{E}) = 0$ (particle-hole symmetry). The five low energy bands correspond to the ``normal'' modes ($| \boldsymbol{\Tilde{u}}(k) | ^2-|\boldsymbol{\Tilde{v}}(k)| ^2 = 1$), while the five upper energy bands correspond to the ``ghost'' modes ($| \boldsymbol{\Tilde{u}}(k) | ^2-|\boldsymbol{\Tilde{v}}(k)| ^2 = -1$). This spectrum highlights that spatially engineering the driving field enables tailoring the Bogoliubov band structure~\cite{Bardyn2016}. Moreover, since the laser is far detuned from the bands ($\Delta E / J \gtrsim 5 $), we note that the ghost and normal branches are well separated in energy, so that their coupling is anticipated to be small.

We now consider a protocol where $\varphi$ is adiabatically varied between $0$ and $2 \pi$, at a rate much smaller than the loss rate $\gamma ^{-1}$. As a result, for any value of $\varphi$, the system reaches a steady-state where drive and loss equilibrate. As we vary $\varphi$, the diagonal elements of the Bogoliubov matrix are periodically modulated according to $\hat{\mathcal{M}}_{n,n} (\varphi) \! = \! E_0 + 2 g  \left| \psi_n ^{(s)} (F_n(\varphi)) \right|^2  - \big( \hbar \omega_{P} + i \frac{\gamma}{2}\big)$. In Fig.~\ref{Fig2}b, we show the evolution of the nonlinear potential $2 g \left| \psi_n ^{(s)} (F_n(\varphi)) \right|^2$ versus $\varphi$. We observe that it defines a series of potential wells that get deformed, in a way that they exhibit a unidirectional motion. 

We point out that $\hat{\mathcal{M}}$ has the form of an Aubry-Andr\'e-Harper Hamiltonian~\cite{Harper1955,Aubry80}, where the onsite energies are modulated as a function of $\varphi$, while the couplings $J$ stay constant. As highlighted in Ref.~\cite{Kraus2012b,Kraus2012}, it is useful to consider the hybrid 1+1D space spanned by real space and the parametric dimension associated with the parameter $\varphi$. In this synthetic-dimension picture, one introduces a 2D operator $\hat{\mathcal{M}}_{\rm 2D}$ defined as:

\begin{equation}
\begin{aligned}
\hat{\mathcal{M}} _{\rm 2D} = \frac{1}{2 \pi} \int_0^{2 \pi} \hat{\mathcal{M}} (\varphi) d \varphi    \ .
\end{aligned}
\label{Eq:M2Dsupmat}
\end{equation}

\noindent We show, in the Supplemental~\cite{SupMat}, that $\hat{\mathcal{M}_{\rm 2D}}$ describes particles hopping on the sites $(n,m)$ of a square lattice with nearest-neighbor coupling $-J$ along the spatial dimension, and complex couplings proportional to $e^{-i 2 \pi \alpha n (m'-m)}$ between sites $(n,m)$ and $(n,m')$ along a synthetic dimension. In the particular case of a pure sine-modulation of the on-site energies, only nearest-neighbor couplings remain along the synthetic dimension. In this situation, $\hat{\mathcal{M}}_{\rm 2D}$ exactly coincides with the Harper-Hofstadter Hamiltonian~\cite{Hofstadter76}, which describes particles experiencing a magnetic field with $2\pi \alpha$ flux per plaquette. We emphasize that time reversal symmetry breaking is provided by the sign of $\alpha$ in the protocol defined in Eq.~{\ref{Eq:protocol}}.

We plot, in Fig.~\ref{Fig3}b, the 1+1D (versus $k$ and $\varphi$) Bogoliubov energy bands outlined during the drive protocol. We observe that all gaps described in Fig.~\ref{Fig2} remain open for all $\varphi$ values. For the five normal and the five ghost branches, we then compute Chern numbers~\cite{Peano2016,footnoteChern} and find $\mathcal{C} = 1, \, 1, \, -4, \, 1, \, 1$. These values correspond to the Chern numbers expected for the Harper-Hofstadter model with magnetic flux $2 \pi \alpha = 2 \pi /5$. These results establish  adiabatic drive modulation of a nonlinear fluid of light as a powerful tool to induce non-trivial topology in an effectively 2D space with one synthetic dimension. Adjusting the drive protocol, one can straightforwardly extend this demonstration to any rational flux $\alpha \! = \! p/q$, where $q$ is the number of sites per unit cell (examples for $\alpha \! = \! 1/3$ and $\alpha \! = \! 1/4$ are provided in the Supplemental~\cite{SupMat}).

\begin{figure}
    \centering
    \includegraphics{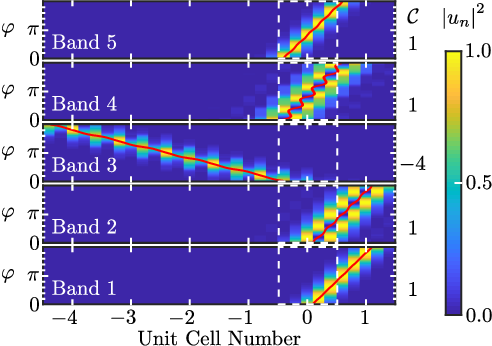}
    \caption{Intensity distributions $\left| u _ n \right| ^2$ of a Wannier state located at a chosen lattice position as a function of $\varphi$. From bottom to top, we represent Wannier states in each of the five normal bands ordered by increasing energies. The red lines trace the barycenter motion of each Wannier state during one protocol period. The white dashed lines delimit one unit cell. The intensity distributions are normalized to their maximum values. All numerical parameters are identical to the ones used in Fig.~3.}
    \label{Fig4}
\end{figure}

To evidence topological pumping for the Bogoliubov excitations, we now investigate the Wannier centers and their evolution during the drive protocol. We compute the maximally-localized Wannier states associated to the five lowest energy bands, and report, in Fig~\ref{Fig4}, their intensity distributions $\left| u _ n \right| ^2$ as a function of $\varphi$. The solid lines follow the barycenter trajectories of the Wannier states, and clearly highlight a net motion. Moreover, after one modulation period, each Wannier state has moved by a quantized number of unit cells that exactly matches the Chern number of the band (see Supplemental~\cite{SupMat} for other values of $\alpha$). This fully demonstrates that we have realized a topological pump for Bogoliubov excitations. The crucial novelty of our result lies in the fact that the pumping mechanism relies entirely on the presence of inter-particle interactions which modulate the potential acting on Bogoliubov excitations.

\begin{figure}
    \centering
    \includegraphics{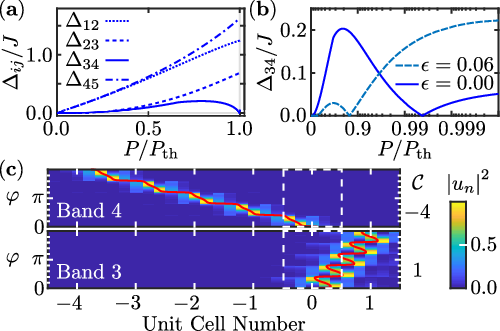}
    \caption{a.~Energy gaps $\Delta_{ij}$ versus $P/P_{\rm th}$. b.~Logarithmic plot of $\Delta_{34}/J$ versus $P/P_{\rm th}$ for $\epsilon = 0$ (solid line) and $\epsilon = 0.06$ (dashed line). c.~Intensity distributions $\left| u _ n \right| ^2$ versus $\varphi$ for the Wannier states associated to the third (bottom) and fourth (top) Bogoliubov bands obtained for $\epsilon=0$, just below $P_{\rm th}$ ($P/P_{\rm th} = 99.99~\%$). The red lines trace the barycenter motion of each Wannier state during one protocol period. The intensity distributions are normalized to their maximum values.}
    \label{Fig5}
\end{figure}

So far, we have focused on a value of $P$ lying well below $P_{\rm th}$. In Figure~\ref{Fig5}a, we plot the evolution of the energy gaps $\Delta_{ij}$ between bands $i$ and $j$, as we increase $P$ up to $P_{\rm th}$. Interestingly, at a critical power $P_c/P_{\rm{th}} = 99.6 \%$, we observe that $\Delta_{34}$ vanishes, and then the gap reopens (see semi-logarithmic plot in Fig.\ref{Fig5}b). For power values within the range ($P_c <  P < P_{\rm{th}} $), we repeat our analysis. We first compute the Chern numbers for the 1+1D Bogoliubov bands, and obtain $\mathcal{C} = 1, \, 1, \, 1, \, -4, \, 1$. Compared to the situation discussed previously in Fig.~\ref{Fig3}b, the Chern numbers of the third and fourth bands are inverted. We then calculate the Wannier centers evolution and find that their barycenter motion follows the new Chern number ordering (see Fig.~\ref{Fig5}c).

These results are a direct consequence of the increasing anharmonicity of the onsite energy modulation as $P$ increases. This is illustrated in Fig.~\ref{Fig2}c, where we compare the populations $\left| \psi_3 ^{(s)} \right|^2$ and the associated power spectra for two different values of $P/P_{\rm th}$. We observe a clear increase of the anharmonic components as $P$ approaches $P_{\rm th}$. As mentioned earlier, the $n$th harmonic of the nonlinear potential introduces nonzero $n$th-neighbor coupling along the synthetic dimension (see Supplemental~\cite{SupMat} for the details). In our protocol, the topological properties of the Harper-Hofstadter band structure stay robust to long-range couplings as long as $P/P_{\rm th} < 99.6~\%$. Above that critical power, the long-range couplings induce a gap closing accompagnied with a topological transition.

Increasing $P$ to values $P>P_{\rm th}$, we observe that the anharmonicity of the onsite energy modulation decreases, and we recover the Harper-Hofstadter Chern ordering. These results illustrate an asset of our pump protocol, whereby the drive amplitude can be used as a knob to control the topology that emerges in the Bogoliubov spectrum. As a further tuning knob, we now show that one can control $P_c$ by modifying the power spectrum of the drive protocol. We illustrate this feature by studying the evolution of $\Delta_{34}$ versus $P/P_{\rm{th}}$ when adding an harmonic of order two to the drive: 
\begin{equation}
F_{n} ( \varphi ) = F  \left| \cos (\varphi/2 + \pi \alpha \, n   ) \right| + \epsilon F  \left| \cos (\varphi + 2 \pi \alpha \, n   ) \right| \, .
\label{Eq:protocolAnharm}
\end{equation}
The result is shown in Fig.~\ref{Fig5}b (dashed line) for $\epsilon = 0.06$. Under the modified drive protocol, the critical power at which the topological transition occurs gets significantly reduced to the value $P^{'}_c = 0.85 P^{'}_{\rm{th}}$.
We note that tuning the number and amplitudes of harmonics included in the drive protocol provides a vast nonlinear optics framework to trigger topological transitions and control the motion of Wannier states in the Bogoliubov excitation spectrum. A systematic study of these effects is beyond the scope of this letter.

In conclusion, starting with a trivial lattice of driven-dissipative Kerr resonators, we establish Bogoliubov excitations as a playground to organize Thouless pumping through spatial engineering of the drive. Adiabatically modulating the drive pattern introduces a parametric dimension leading to 1+1D topological bands. As a consequence of the presence of non-zero Chern numbers, the Bogoliubov Wannier states associated to the bands experience quantized motion through Thouless pumping. Moreover, the nonlinearity modifies the power spectrum of the onsite energy modulation, leading to topological transitions that switch the directionality and amplitude of the quantized motion of Wannier states within different bands. Our simulations are performed using realistic parameters that are compatible with the exciton polariton platform. Benefiting from recent progress in probing and manipulating Bogoliubov excitations in these systems~\cite{Stepanov2019,Claude2022,Frerot2023}, we foresee that the experimental implementation of the proposed topological pump is within reach. As a further development, we envision that drive engineering can be used to induce uniform magnetic flux for Bogoliubov excitations in 2D lattices. By extending the perturbative approach of Ref.~\cite{Bardyn2016}, one could realize Hofstadter-type models with large synthetic flux $\Phi\sim\pi$ per plaquette, hence offering new opportunities to explore nonlinear topology in 2D lattices~\cite{tesfaye2024,villa2024}.

\begin{acknowledgments}
We thank A.~Amo, I.~Carusotto, and O.~Zilberberg for fruitful discussions. This work was partly supported by the Paris Ile de France R\'egion in the framework of DIM SIRTEQ, by the European Research Council (ERC) under the European Union's Horizon 2020 research and innovation programme (project ARQADIA, grant agreement no.~949730), and under Horizon Europe research and innovation programme (ANAPOLIS, grant agreement no.~101054448). NG is supported by the ERC Grant LATIS and the EOS project CHEQS. NM acknowledges funding by the Deutsche Forschungsgemeinschaft (DFG, German Research Foundation) under Germany's Excellence Strategy -- EXC-2111 -- 390814868.
\end{acknowledgments}

\clearpage

\onecolumngrid

\section*{Supplemental material for: \\ Thouless pumping in a driven-dissipative Kerr resonator array}

In this Supplemental, we provide additional information and data regarding our work on Thouless pumping in a driven-dissipative Kerr resonator array. Firstly, we detail the derivation of the mapping of our model to the Harper-Hofstadter Hamiltonian. Secondly, we show additional data for $\alpha =1/3$ and $\alpha = 1/4$, illustrating the generality of the proposed Thouless pumping scheme.

\section{Mapping to the Harper-Hofstadter model}

We have shown in Eq.~(4) of the main text, that the Bogoliubov excitations in a driven system of coupled Kerr resonators are described by the following matrix:

\begin{equation}
\begin{aligned}
\hat{\mathcal{B}} =
\begin{pmatrix}
\hat{\mathcal{M}} & \hat{\mathcal{N}} \\
-\hat{\mathcal{N}}^{*} & -\hat{\mathcal{M}}^{*}
\end{pmatrix}
 \, .
\end{aligned}
\label{SupEq:BogoMat}
\end{equation}

\noindent By construction of the Bogoliubov matrix, $\hat{\mathcal{B}}$ is twice the size of the system, and can be conveniently divided it into four distinct blocks. The top-left sector $\hat{\mathcal{M}}$ is associated to the normal modes, while the bottom right block $-\hat{\mathcal{M}}^*$ relates to the ghost modes. As for the non-diagonal blocks $\hat{\mathcal{N}}$ and $\hat{-\mathcal{N}^*}$, they describe the coupling between the normal and ghost modes. In this Letter, we consider a pump energy detunings that are large compared to the amplitude of the (linear) energy bands ($\Delta E / J \gg 1$). Therefore, the normal and ghost modes are far detuned from each other and they do not mix significantly. Focusing on the physics of the normal modes, one can thus restrict to the study of $\hat{\mathcal{M}}$.

\,

Using Eq.~(1) of the main text, $\hat{\mathcal{M}}$ can be expressed as:

\begin{equation}
\begin{aligned}
\hat{\mathcal{M}} (\varphi) = \sum_{n} \bigg( E_0 - \hbar \omega_{\rm P} + 2 g N_{n}^{(s)} (\varphi) - i\frac{\gamma}{2} \bigg) \dyad{u_n(\varphi)}{u_n(\varphi)} - J \bigg( \dyad{u_n(\varphi)}{u_{n+1}(\varphi)} + \dyad{u_n(\varphi)}{u_{n-1}(\varphi)} \bigg)  \ ,
\end{aligned}
\label{Eq:Msupmat}
\end{equation}

\noindent where $N_{n}^{(s)}$ is the stationary number of particles at site $n$. As mentionned in the main text, $\hat{\mathcal{M}} (\varphi)$ has the general form of a 1D Aubry-Andr\'e-Harper Hamiltonian, where the onsite energy of site $n$ depends on $\varphi$, while the hopping amplitude between neighboring sites is independent of $\varphi$. We emphasize that $\varphi$ varies adiabatically, on time scales that are much slower that the inverse particle linewidth $\gamma^{-1}$, so that a steady-state is reached at each instant of the protocol. Moreover, as highlighted in Fig.~2(b), $N_{n}^{(s)}(\varphi)$ depends periodically on $\varphi$ (with period $2\pi$) and on $n$ (with period $\alpha^{-1}$). Therefore, it can be expanded into a Fourier series as follows:

\begin{equation}
\begin{aligned}
N_{n}^{(s)}(\varphi) = \sum_{l=- \infty}^{\infty} C_l e^{i l ( 2 \pi \alpha n + \varphi )}   \ .
\end{aligned}
\label{Eq:FourierSupmat}
\end{equation}

\,

Following the approach developed in Ref.~[35] and Ref.~[3] of the main text, we now show that the parameter $\varphi$ can be seen as a parametric dimension, and $\hat{\mathcal{M}} (\varphi)$ can be mapped to a 2D system. In brief, each realization of $\hat{\mathcal{M}} (\varphi)$ can be taken as the Fourier transform of a 2D operator $\hat{\mathcal{M}} _{\rm 2D}$ along an extended dimension, defined as:

\begin{equation}
\begin{aligned}
\hat{\mathcal{M}} _{\rm 2D} = \frac{1}{2 \pi} \int_0^{2 \pi} \hat{\mathcal{M}} (\varphi) d \varphi    \ .
\end{aligned}
\label{Eq:M2Dsupmat}
\end{equation}

\noindent We then decompose $\ket{u_n(\varphi)}$ into the Fourier space associated to $\varphi$ as follows:

\begin{equation}
\begin{aligned}
\ket{u_n(\varphi)} = \sum_m e^{  i \varphi m} \ket{\mathcal{U}_{n,m}}   \ .\\
\end{aligned}
\end{equation}

\noindent In the following, we note $\ket{\mathcal{U}_{n,m}} = \ket{n,m}$. Inserting these expressions into $\hat{\mathcal{M}} _{\rm 2D}$, one gets:

\begin{equation}
\begin{split}
\hat{\mathcal{M}} _{\rm 2D} & =  \sum_n \left( E_0 - \hbar \omega_{\rm P} - i \frac{\gamma}{2} \right) \frac{1}{2\pi} \int_{0}^{2 \pi} \sum_{m,m'} e^{i \varphi (m - m')} d \varphi \dyad{n,m}{n,m'} \\
& + 2 g \sum_{n} \sum_{l=-\infty}^{\infty} C_l e^{i l 2 \pi \alpha n } \frac{1}{2\pi} \int_0^{2 \pi} \sum_{m,m'}  e^{i (l+m-m') \varphi} d \varphi \dyad{n,m}{n,m'} \\
& -J \sum_n \frac{1}{2 \pi} \int_0^{2 \pi} \sum_{m,m'} e^{i \varphi (m - m')} d \varphi \dyad{n,m}{n+1,m'} \\
& -J \sum_n \frac{1}{2 \pi} \int_0^{2 \pi} \sum_{m,m'} e^{i \varphi (m - m')} d \varphi \dyad{n,m}{n-1,m'}
\ .
\end{split}
\end{equation}

\noindent Therefore, we map $\hat{\mathcal{M}} (\varphi)$ to the following 2D operator:

\begin{equation}
\begin{split}
\hat{\mathcal{M}} _{\rm 2D} &=  \sum_{n,m} \, \Big \{  \left( E_0 - \hbar \omega_{\rm P} - i \frac{\gamma}{2} \right) \dyad{n,m}{n,m}
-J  \left( \dyad{n,m}{n+1,m} + \dyad{n,m}{n-1,m} \right) \\
& + 2 g \sum_{l=-\infty}^{\infty} C_l e^{i l 2 \pi \alpha n } \dyad{n,m}{n,m+l} \Big \} \, .
\end{split}
\label{M2Dv0}
\end{equation}

\begin{figure}
    \centering
    \includegraphics{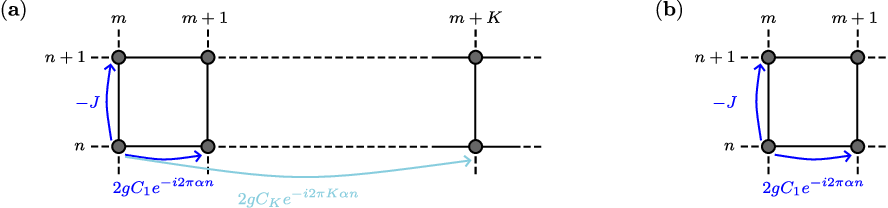}
    \caption{Mapping $\hat{\mathcal{M}}(\varphi)$ to a 2D model describing particle hopping on a 2D square lattice by the Hamiltonian $\hat{\mathcal{M}}_{\mathrm{2D}}$. The $\varphi$-dependence of $\hat{\mathcal{M}}(\varphi)$ is then understood as the Fourier transform of $\hat{\mathcal{M}}_{\mathrm{2D}}$ along the extended dimension labeled by $m$. a.~Most general case, where the particles experience real hopping $-J$ along one direction and $l^{\rm th}$-neighbor complex hoping $\propto e^{-i 2 \pi l \alpha n}$ along the other direction.  b.~When the onsite energy modulation is a purely harmonic function, one recovers the Harper-Hofstadter model. The particles experience a $2\pi \alpha$ magnetic flux per plaquette.}
    \label{FigS1}
\end{figure}

\noindent The operator $\hat{\mathcal{M}} _{\rm 2D}$ describes the physics of non-interacting particles on a 2D lattice. The particles experience nearest neighbor coupling along the spatial dimension, with hopping coefficient equal to $-J$. Along the orthogonal synthetic direction, the particles experience $l^{\rm th}$-neighbor couplings from site $(n,m)$ to site $(n,m+l)$, with complex hopping coefficients equal to $2g C_l e^{i l 2 \pi \alpha n}$ (see Fig.~\ref{FigS1}(a)).

\,

We note that the long-range couplings in $\hat{\mathcal{M}} _{\rm 2D}$ are introduced by the higher harmonics appearing in the Fourier decomposition of $N_n^{(s)}(\varphi)$, as a consequence of the Kerr nonlinearity. In the limit where $N_n^{(s)}(\varphi)$ is a harmonic function ($C_l = C_1 \delta_{1l}$), $\hat{\mathcal{M}} _{\rm 2D}$ writes:

\begin{equation}
\begin{split}
\hat{\mathcal{M}} _{\rm 2D} =  & \sum_{n,m} \left( E_0 - \hbar \omega_{\rm P} + 2gC_0 - i \frac{\gamma}{2} \right) \dyad{n,m}{n,m}
-J  \left( \dyad{n,m}{n+1,m} + \dyad{n,m}{n-1,m} \right) \\
& + 2 g C_1 \left( e^{i 2 \pi \alpha n } \dyad{n,m}{n,m+1} + e^{- i 2 \pi \alpha n } \dyad{n,m}{n,m-1} \right)
\ .
\end{split}
\label{Eq:M2Dv1}
\end{equation}

\noindent This corresponds exactly to a Harper-Hofstadter model, with hopping coefficients as shown in Fig.~\ref{FigS1}(b). In particular, we observe the emergence of a $2 \pi \alpha$ magnetic flux per plaquette. We emphasize that, in the 2D model, time-reversal symmetry arises as a consequence of our choice of $\alpha$, which determines the phase difference between neighboring drives in the pumping protocol (see Eq.~(3) of the main text). The effective magnetic field can thus be inverted by changing the sign of $\alpha$.

\,

\begin{figure}[h!]
    \centering
    \includegraphics{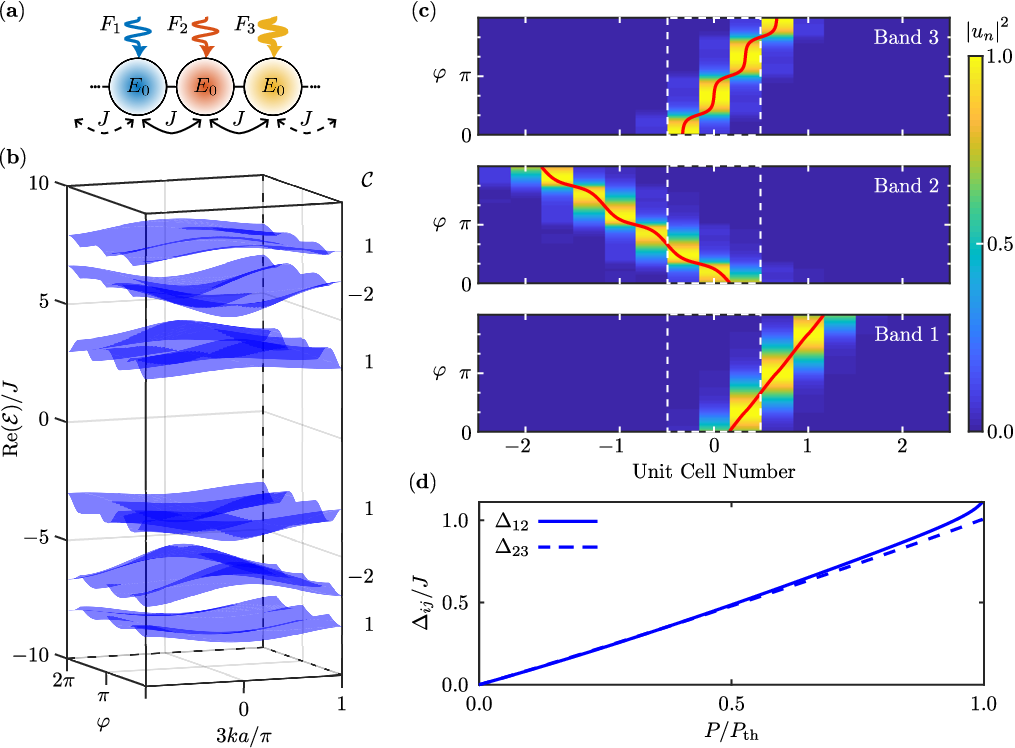}
    \caption{The $\alpha \! = \! 1/3$ case. a.~Schematic of the lattice of coupled Kerr resonators, with three sites per unit cell. b.~1+1D Bogoliubov bands as a function of $k$ and $\varphi$. The numerical parameters are the same as in the main text. $P / P_{th} = 0.91$. c.~Intensity distributions of the Wannier states in each (normal) band versus lattice position and $\varphi$. From bottom to top, the three normal bands are ordered by increasing energies. The red lines trace the barycenter motion of each Wannier state during one protocol period. The white dashed lines delimit one unit cell. The intensity distributions are normalized to their maximum values. d.~Energy gaps $\Delta_{ij}$ versus $P/P_{\rm th}$.}
    \label{FigS2}
\end{figure}

\section{Additional data for other (rational) values of $\alpha$ }

The drive protocol described in Eq.~(3) of the main text expresses the distribution of the pump amplitudes in the most general way as:

\begin{equation}
    F_{n} (\varphi) = F \left| {\rm cos} (\varphi / 2 + \pi \alpha n) \right| \, ,
\end{equation}

\noindent where $\alpha$ is a parameter encoding the spatial distribution of the drive amplitudes. For rational values $\alpha = p/q$, the drive pattern is spatially periodic, with $q$ sites per unit cell. For irrational values, $\alpha$ is incommensurate with the lattice discretization, and the drive patterns becomes aperiodic. In this Letter, we focus on the canonical version of Thouless pumping, which requires a periodic drive pattern. We thus concentrate on rational values of $\alpha$ and, without loss of generality, consider $\alpha = 1/q$.

\,

As developed above, the system maps to a 2D Harper-Hofstadter model $2 \pi \alpha$ magnetic flux per unit cell. From general considerations about the Harper-Hofstadter model, we expect the spectrum of the system to feature $q$ bands with the following properties:

\begin{enumerate}
    \item If $q$ is an odd number, all the bands are well separated. The middle band Chern number equals $\mathcal{C} = -q+1$. The Chern number for all the other bands is $\mathcal{C}=1$.
    \item If $q$ is an even number, the two middle bands are not gapped and show Dirac touching points. All the other bands are well separated and show a Chern number $\mathcal{C}=1$. The Chern number for the middle pair of bands is $\mathcal{C} = -q+2$. 
\end{enumerate}

\noindent In the main text of this Letter, we focus on $\alpha = 1/5$, where all the Harper-Hofstadter bands are expected to be gaped and well separated in energy ($\alpha = 1/q$ with $q$ odd integer). Additionally, as shown in Fig.~5 of the main text, the $\alpha = 1/5$ case features an interesting topological transition when the power increases, highlighting a new degree of control over Thouless pumps that can be achieved through the nonlinearity. In this Supplemental, we show the results obtained for other values of $\alpha$, thus illustrating the generality of our approach.

\,

We show, in Fig.~\ref{FigS2}, the results obtained for $\alpha \! = \! 1/3$. In this pumping configuration, the unit cell contains three sites, as shown in Fig.~\ref{FigS2}(a). Keeping the same numerical parameters as in the main text, we calculate the Bogoliubov energy spectrum as a function of $k$ and $\varphi$ for a pump power $P / P_{\rm th} = 0. 91$. The resulting spectrum is plotted in Fig.~\ref{FigS2}(b), where we observe the presence of six Bogoliubov bands, three normal bands and three ghost bands. As anticipated from the mapping to the Harper-Hofstadter model, the bands are gaped and well separated in energy, and feature non-zero Chern numbers equal to 1, -2 and 1 from the lowest energy (normal) band to the highest energy (normal) band. For each normal band, we then calculate the Wannier state evolution during one period of the pump protocol. We observe, in Fig.~\ref{FigS2}(c) that the Wannier states experience a net motion. Moreover, their motion is quantized and matches the Chern numbers of the bands, as expected for a Thouless pump. Finally, we show, in Fig.~\ref{FigS2}(d) the evolution of the energy gaps $\Delta_{12}$ and $\Delta_{34}$ versus $P/P_{\rm th}$. We observe that the size of these gaps steadily increases and reaches sizeable values $\Delta / J \simeq 1$ as $P$ reaches $P_{\rm th}$. Contrary to what was observed in the $\alpha = 1/5$ case, no gap closure, and thus no topological transition is observed here. Indeed the anharmonicities induced by the nonlinearity are too weak to overcome the size of the gaps, and the system stays in the Harper-Hofstadter configuration for all values of $P / P_{\rm th}$.

\,

\begin{figure}
    \centering
    \includegraphics{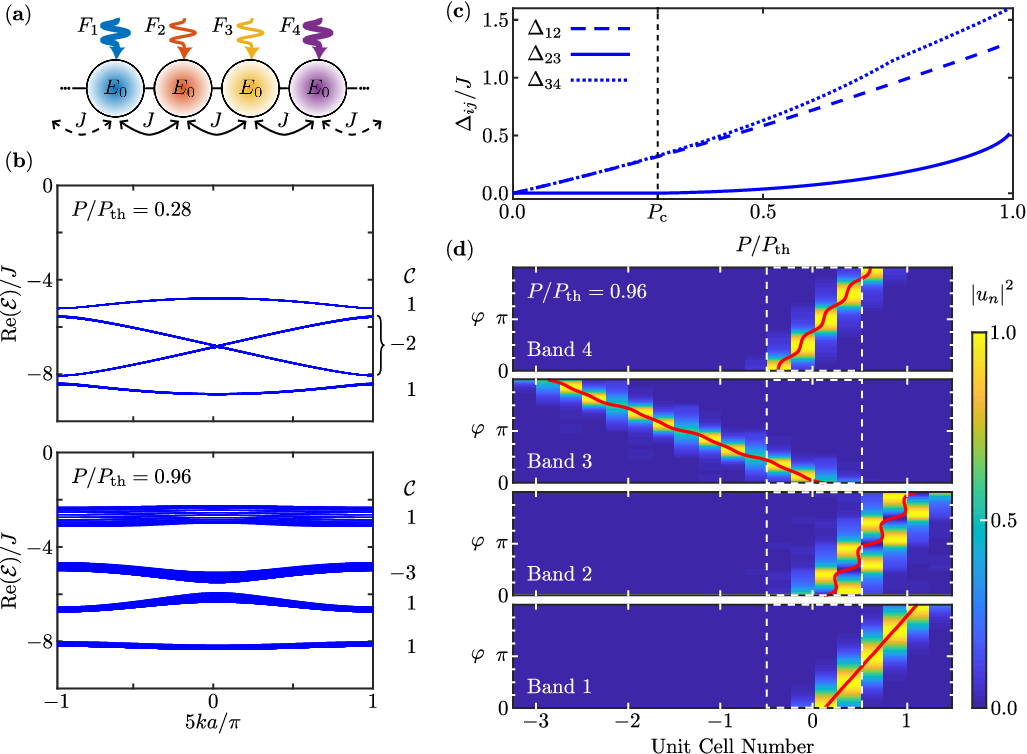}
    \caption{The $\alpha \! = \! 1/4$ case. a.~Schematic of the lattice of coupled Kerr resonators, with four sites per unit cell. b.~The Bogoliubov (normal) bands as a function of $k$, plotted for all values of $\varphi$. All numerical parameters are the same as in the main text. The top (bottom) panel shows the bands for $P / P_{th} = 0.28$ ($P / P_{th} = 0.96$). c.~Energy gaps $\Delta_{ij}$ versus $P/P_{\rm th}$. d.~Intensity distributions of the Wannier states calculated in each (normal) band versus lattice position and $\varphi$ for $P / P_{th} = 0.96$. From bottom to top, the four bands are ordered by increasing energies. The red lines trace the barycenter motion of each Wannier state during one protocol period. The white dashed lines delimit one unit cell. The intensity distributions are normalized to their maximum values.}
    \label{FigS3}
\end{figure}

Finally, we show in Fig.~\ref{FigS3} the results obtained for $\alpha = 1/4$, where the pumping protocol contains four sites per unit cell (see Fig.~\ref{FigS3}(a)). We plot, in Fig.~\ref{FigS3}(b), the normal bands of the Bogoliubov spectrum for two different values of $P/P_{\rm th}$. For a pump power well below $P_{\rm th}$ ($P/ P_{\rm th} = 0.28$, bottom panel), we observe that the two outermost bands of the spectrum are gaped from the other bands, while the two middle bands are not gaped and feature a Dirac touching point at $k=0$. Calculating the Chern numbers for these bands, we obtain $\mathcal{C} = 1$ for the two outermost bands, while $\mathcal{C} = -2$ for the middle pair of bands. These results are in perfect agreement with what is expected from the mapping to the 2D Harper Hofstadter Hamiltonian. On the contrary, for $P/ P_{\rm th} = 0.96$ (top panel), we observe that the two middle bands are now well separated, and the Chern numbers of the four bands are $\mathcal{C} = 1, ~1, ~ -3, ~1$, in increasing order of band energies. To further investigate this effect, we plot, in Fig.~\ref{FigS3}(c), the evolution of the gap energies $\Delta_{ij}$ versus $P/P_{\rm th}$. Interestingly, we notice that for a range $ 0 \leq P/P_{\rm th} < P_{\rm c} = 0.295$, the middle gap energy $\Delta_{23}$ stays locked to zero. Above this critical value, the two middle bands no longer feature a Dirac touching point ($\Delta_{23} > 0$). This middle gap opening is a direct consequence of the system nonlinearity. Indeed, for increasing values of $P/P_{\rm th}$, the anharmonic content in the modulation of the onsite energies introduces long range couplings in $\mathcal{M} _ {\rm 2D}$ (see Eq.~\ref{M2Dv0}), leading to the emergence of physical features that go beyond the Harper-Hofstadter model when these long range couplings become significant ($P/P_{\rm th} > P_{\rm c}$). In this regime ($P/P_{\rm th} = 0.96$), we calculate the evolution versus $\varphi$ of the Wannier states associated to each (normal) band during one period of the pump protocol. We observe, in Fig.~\ref{FigS3}(d) that the net motion experienced by the Wannier states is quantized and matches the Chern numbers calculated for the
bands. This again confirms that the systems realizes a Thouless pump.

\clearpage
\twocolumngrid


%

\end{document}